\input harvmac

\def\npb#1#2#3{{\it Nucl.\ Phys.} {\bf B#1} (19#2) #3}

\def\prdm#1#2#3{{\it Phys.\ Rev.} {\bf D#1} (20#2) #3}

\def\h{{1 \over 2}}
\def\frac#1#2{{#1 \over #2}}
\def\ith#1{\Theta^{-1}_{#1}}
\def\CH{{\cal H}}
\def\CO{{\cal O}}
\def\CS{{\cal S}}
\def\dD{{\bar D}}
\def\bphi{{\bar \phi}}
\def\bS{{\bar S}}
\def\bz{{\bar z}}
\def\hF{{\widehat{F}}}
\def\II{{I\!I}}

\def\dphi{{\delta\phi}}
\def\vp{{\varphi}}
\def\bbeta{{\bar \beta}}
\lref\sena{A.~Sen, ``Non-BPS States and Branes in String Theory,''  
hep-th/9904207.}
\lref\senb{A.~Sen, ``Descent Relations Among Bosonic D-Branes,'' {\it
Int.\ J.\ Mod.\ Phys.} {\bf A14} (1999) 4061, hep-th/9902105.}
\lref\senc{A.~Sen, ``SO(32) Spinors of Type I and Other Solitons on
Brane-Antibrane Pair,'' JHEP {\bf 9809} (1998) 023, hep-th/9808141.}
\lref\bosref{J.~A.~Harvey and P.~Kraus, ``D-branes as Unstable Lumps in
Bosonic Open String Field Theory,'' JHEP {\bf 0004} (2000) 012,
hep-th/0002117; R.~de Mello Koch, A.~Jevicki, M.~Mihailescu, and R.~Tatar,
``Lumps and p-Branes in Open String Field Theory,'' {\it Phys.\ Lett.}
{\bf B482} (2000) 249, hep-th/0003031; N.~Moeller, A.~Sen, and B.~Zwiebach,
``D-branes as Tachyon Lumps in String Field Theory,'' JHEP {\bf 0008},
(2000) 039, hep-th/0005036.}
\lref\leru{A.~Lerda and R.~Russo, ``Stable Non-BPS States in String
Theory: A Pedagogical Review,'' {\it Int.\ J.\ Mod.\ Phys.} {\bf A15}
(2000) 771, hep-th/9905006.}
\lref\bega{O.~Bergman and M.~Gaberdiel, ``Non-BPS Dirichlet Branes,''
{\it Class.\ Quant.\ Grav.} {\bf 17} (2000) 961, hep-th/9908126.}
\lref\cds{A.~Connes, M.~Douglas and A.~Schwarz, ``Noncommutative Geometry
and Matrix Theory: Compactification on Tori,'' JHEP {\bf 02} (1998) 003;
hep-th/9711162.}
\lref\schom{V.~Schomerus, ``D-Branes and Deformation Quantization'',
JHEP {\bf 9906} (1999) 030, hep-th/9903205.}
\lref\witt{E.~Witten, ``Noncommutative Tachyons and String
Field Theory,'' hep-th/0006071.}
\lref\witk{E.~Witten, ``D-Branes and K-Theory,'' JHEP {\bf 9812} (1998)
019, hep-th/9810188.} 
\lref\sw{N.~Seiberg and E.~Witten, ``String Theory and Noncommutative
Geometry,'' JHEP {\bf 9909} (1999) 032, hep-th/9908142.}
\lref\abo{A.~Abouelsaood, C.~G.~Callan, C.~R.~Nappi, and S.~A.~Yost,
``Open Strings In Background Gauge Fields,'' \npb{280}{87}{599}.}
\lref\gava{E.~Gava, K.~S.~Narain, and M.~H.~Sarmadi, ``On the Bound
States of p- and (p+2)-Branes,'' \npb{504}{97}{214}, hep-th/9704006.}
\lref\gms{R.~Gopakumar, S.~Minwalla, and A.~Strominger, ``Noncommutative
Solitons,'' JHEP {\bf 0005} (2000) 020, hep-th/0003160.}
\lref\gmss{R.~Gopakumar, S.~Minwalla, and A.~Strominger, ``Symmetry
Restoration and Tachyon Condensation in Open String Theory,'' JHEP {\bf
0104} (2001) 018, hep-th/0007226.}
\lref\agms{M.~Aganagic, R.~Gopakumar, S.~Minwalla, and A.~Strominger,
``Unstable Solitons in Noncommutative Gauge Theory,'' hep-th/0009142.}
\lref\hkl{J.~A.~Harvey, P.~Kraus, and  F.~Larsen, ``Exact Noncommutative
Solitons,'' JHEP {\bf 0012} (2000) 024, hep-th/0010060.}
\lref\hkle{J.~A.~Harvey, P.~Kraus, and  F.~Larsen, ``Tensionless Branes
and Discrete Gauge Symmetry,'' \prdm{63}{01}{026002}, hep-th/0008064.}
\lref\hklm{J.~A.~Harvey, P.~Kraus, F.~Larsen, and E.~J.~Martinec,
``D-Branes and Strings as Non-commutative Solitons,''
JHEP {\bf 0007} (2000) 042, hep-th/0005031.}
\lref\sei{N.~Seiberg, ``A Note on Background Independence in Noncommutative
Gauge Theories, Matrix Model and Tachyon Condensation,'' JHEP {\bf 0009} (2000)
003, hep-th/0008013.}
\lref\deAl{S.~P.~de Alwis and A.~T.~Flournoy, ``Some Issues in
Noncommutative Solitons as D-branes,'' \prdm{63}{01}{106001}, hep-th/0011223.}
\lref\atsey{A.~A.~Tseytlin, ``On Non-Abelian Generalisation of
Born-Infeld Action in String Theory,'' \npb{501}{97}{41},
hep-th/9701125; A.~A.~Tseytlin, ``Born-Infeld Action, Supersymmetry and
String Theory,'' hep-th/9908105.}
\lref\dmr{K.~Dasgupta, S.~Mukhi, and G.~Rajesh, ``Noncommutative Tachyons,''
JHEP {\bf 0006}, (2000) 022, hep-th/0005006.}
\lref\sak{J.~J.~Sakurai, {\it Modern Quantum Mechans}, revised edition,
New York: Addison-Wesley (1994).}
\lref\cole{S.~Coleman, {\it Aspects of Symmetry}, New York: Cambridge
University Press (1985), Chapter 7.6.}
\lref\kl{P.~Kraus and F.~Larsen, ``Boundary String Field Theory of the
$D\dD$ System,'' \prdm{63}{01}{106004}, hep-th/0012198.}
\lref\ttu{T.~Takayanagi, S.~Terashima, and T.~Uesugi, ``Brane-Antibrane
Action from Boundary String Field Theory,'' JHEP {\bf 0103} (2001) 019,
hep-th/0012210.} 
\Title{\vbox{\baselineskip12pt
\hbox{hep-th/0101125}
\hbox{EFI-2000-56}
\vskip-.5in}}
{\vbox{\centerline{Noncommutative Solitons and Intersecting D-Branes}}}
\medskip\bigskip
\centerline{Li-Sheng Tseng}
\bigskip\medskip
\centerline{\it Enrico Fermi Institute and Department of Physics}
\centerline{\it  University of Chicago,} 
\centerline{\it  5640 S. Ellis Ave., 
Chicago, IL 60637, USA}
\centerline{\tt lstseng@theory.uchicago.edu}
\medskip
\baselineskip18pt
\medskip\bigskip\medskip\bigskip\medskip
\baselineskip16pt
\noindent
We construct intersecting D-branes as noncommutative solitons in bosonic
and type II string theory.  ``Defect'' branes, which are D-branes
containing bubbles of the closed string vacuum, play an important role in
the construction.

\Date{December, 2000}

\newsec{Introduction}

Brane configurations that contain a tachyonic mode are unstable and
will decay.  The decay modes vary depending on the unstable $D$-brane
configurations involved (see the reviews \refs{\sena, \leru, \bega}).  In
bosonic string theory, all $D$-branes are unstable.  The tachyon on the
brane may condense to solitonic solutions of the open string tachyon
field theory.  These solitons have been noted to correspond to lower
dimensional unstable $D$-branes \refs{\senb, \bosref}.  In type II
superstrings, there are stable tachyon vortex solutions arising from
$D\dD$ brane annihilation \refs{\senc, \witk} corresponding to
codimension two stable BPS $D$-branes.

It was noticed \refs{\dmr, \hklm} that the construction of solitonic
$D$-branes solutions is remarkably simplified in the presence of an infinite
non-zero B-field.  The tachyon field theory becomes noncommutative and
using the operator formalism, the $D$-brane solutions, now
noncommutative (NC) solitons \gms, are easily constructed.  These NC
solitonic $D$-branes solutions have since been improved to allow for
arbitrary non-zero finite B-field \hkl .  

In this paper, we present new decay modes of unstable brane
configurations in the presence of non-zero B-field.   Our aim is to
construct decay modes within the context of the NC operator formalism
corresponding to intersecting branes configurations.  The intersecting
branes we construct all have a non-zero B-field in their worldvolume.

In section 2, we carry this out within the context of bosonic string theory,
using as a specific example the decay of unstable $D25$-branes.  We find
that configurations of two intersecting $D(p-2)$-branes can be
constructed only from the decay of two or more $Dp$-branes.  Starting from
one $Dp$-brane, one can obtain perpendicular brane configurations
consisting of $D(p-4)$-branes plus defect $D(p-2)$-branes.  Defect
branes are D-branes containing bubbles of the closed string vacuum.
The defect $Dp$-brane soliton is time-independent and asymptotically
approaches the solution of the $Dp$-brane.   We conclude the section by
constructing non-perpendicularly intersecting $D$-branes.

In section 3, we extend our results to the type II
superstrings.  Special to the superstring case is that perpendicularly
intersecting brane configurations can be constructed from the decay of a
single $Dp$-$\dD p$ system.  In particular, it is possible to obtain a
$D(p-2)$-brane perpendicularly intersecting a $\dD(p-2)$-brane from the
decay of a single $Dp$-$\dD p$ brane configuration. 

In labeling coordinates, we will let $x^{i}$ denote the noncommutative
directions, $x^{a}$ the commutative directions, and $x^{\mu}$
all the directions.

\newsec{$D25$-Branes Decay Modes}

\subsec{Action and Conventions}

The effective action for the tachyon field, $\phi$, and the U(1) gauge field in
the presence of a non-zero B-field in Euclidean space is    
\eqn\ac{
\CS={c \over g_s} \int\! d^{26}x
\left\{V(\phi-1)\sqrt{\det(g+2\pi\alpha'(B+F))}+{\alpha' \over
2}\sqrt{g}f(\phi-1)g^{\mu\nu}\partial_{\mu}\phi\partial_{\nu}\phi+ \cdots \right\}.}
where we have left out higher derivative terms.  $V$ and $f$
are written as functions of $\phi-1$.  $\phi$ has been defined such that
at the local maximum $\phi=0$, $V(-1)=1$, and at the local minimum $\phi=1$,
$V(0)=f(0)=0$.  Thus, the tension of a $D25$-brane with zero B- and
F-field
is $T_{25}=c/g_{s}$.  With a non-zero $B$, the tension
contains an extra factor, $T'_{25}=T_{25}\sqrt{\det(g+2\pi\alpha'B)}$.

As shown in \refs{\schom, \sw}, the presence of a constant non-zero
B-field can also be incorporated in the effective action by making the
space coordinates noncommutative, i.e. $[x^{i},x^{j}]=i\Theta^{ij}$ with
ordinary products replaced by $*$ products.  In this
description, the closed string metric and coupling are replaced by the
open string metric $G_{ij}$ and coupling $G_{s}$.  The gauge
symmetry becomes the noncommutative U(1) with the tachyon transforming
in the adjoint representation and the noncommutative field strength
$\hF$ appearing only in the combination $\hF+\Phi$, where $\Phi$ is an
additional parameter.  Making the choice $\Phi=-B$, the relations between
closed and open string variables for a maximum rank B-field are
given in Euclidean signature by 
\eqn\swmap{\eqalign{\Theta=& {1 \over B}~, \cr
G =& -(2\pi \alpha')^2 B{1 \over g}B~, \cr
G_s =&~ g_s \det(2\pi\alpha' Bg^{-1})^\h~.}}  
Introducing $C_{i}=\ith{ij}x^{j}+A_{i}$ and recalling that the derivative
is defined in NC theories as  $\partial_{i}f=-i\ith{ij}[x^{j},f]$, we
write
\eqn\phiF{\eqalign{D_{i} \phi=& \partial_{i} \phi
-i[A_{i},\phi]=-i[C_{i},\phi]~, \cr
\hF_{ij}+\Phi_{ij}=&
\partial_{i} A_{j}-\partial_{j} A_{i}-i[A_{i},A_{j}]-B_{ij}=-i[C_{i},C_{j}]~.}}
Therefore, the Euclidean noncommutative effective action for a maximum rank
B-field can be written as
\eqn\ncac{\eqalign{
\CS={c \over G_s} \int\! d^{26}x
&\left \{ V(\phi-1)\sqrt{\det(G_{ij}-2\pi\alpha'i[C_{i},C_{j}])}\right. \cr
&~~~-\left.{\alpha' \over
2}\sqrt{G}f(\phi-1)G^{ij}[C_{i},\phi][C_{j},\phi]+ \cdots \right \}~,}}
with equations of motion
\eqn\nceqm{\eqalign{\delta\phi:&\qquad
V'(\phi-1)\sqrt{M}-\frac{\alpha'}{2}\sqrt{G}G^{ij}\left\{f'(\phi-1)[C_{i},\phi][C_{j},\phi]+2[f(\phi-1)[C_{i},\phi],C_{j}]\right\}=0
\cr
\delta C_{i}:&\qquad
-2\pi i[C_{j},V(\phi-1)\sqrt{M}(M^{-1})^{ji}]-\sqrt{G}G^{ij}[\phi,f(\phi-1)[C_{j},\phi]]=0}}
where $M_{ij}=G_{ij}-2\pi\alpha'i[C_{i},C_{j}]$.

Although we will carry out calculations throughout in Minkowski
space and with a non-maximum rank B-field, \ncac\ times an overall minus
sign is the action we will use.  This is appropriate because all solitonic
solutions presented in this paper are independent of the commutative
coordinates and have vanishing $A_{a}$, the commutative components of
the gauge field.  All solutions are also exact \refs{\gmss,\hkl} in that
each satisfies \nceqm\ and also $[C_{i},\phi]=0$ and
$[C_{i},[C_{j},C_{k}]]=0$.  The latter two conditions ensures that
higher derivative terms in the action \ncac\ do not affect the
equations of motion and the energy of the solution.

Working in the NC operator formalism, fields on noncommutative space
become operators acting on an auxiliary infinite-dimensional Hilbert
space with dependence only on the commutative coordinates.  With a
non-zero B-field in only two directions, the Hilbert space is $\CH=L^2(R)$
and we will use as a basis the eigenstates of the simple harmonic
oscillator.  The projection operator $P_n$ and the shift operator $S_n$ are defined for $n \geq 1$ to be
\eqn\pshift{P_{n}=\sum_{m=0}^{n-1}|m\rangle \langle m|~,\qquad
S_{n}=\sum_{m=0}^{\infty}|m+n\rangle \langle m|~,}
satisfying the relations
\eqn\shiftrel{S_{n_1}S_{n_2}=S_{n_1+n_2}~,\qquad \bS_{n} S_{n} = I~,\qquad S_{n}\bS_{n}=I-P_{n}~,}
\eqn\pshiftrel{P_{n}S_{n}=\bS_{n}P_{n}=0~.}
For $\Theta^{ij}$ of rank $2p$, the corresponding Hilbert space can be
expressed as a tensor product of $p$ copies of $\CH$.  As an example,
for $p=2$, operators act on the Hilbert space $\CH\otimes\CH$.
Projection operators on such a space are likewise obtained by tensor
products, for example
\eqn\genproj{P^{I}_{n_1}P^{\II}_{n_2}\equiv P_{n_1}\otimes P_{n_2}=\sum_{m_1=0}^{n_1-1}\sum_{m_2=0}^{n_2-1}|m_{1},m_{2}\rangle\langle m_{1},m_{2}|~.}
Our notation is that the Roman number superscript on the operator will refer
to the Hilbert space on which it acts.  The superscript is left out when
there is no ambiguity.  For instance, $P$ and $S$ are taken to be,
respectively, any general projection and shift operator acting on the
whole auxiliary Hilbert space.  Finally, integrals over noncommutative
space are expressed in the operator formalism as
\eqn\intr{\int \frac{d^{2p}x}{\sqrt{\det(\Theta^{ij})}} \to (2\pi)^{p}{\rm Tr}~.}

\subsec{$D21$-branes vs. $D23$-brane with Tangential B-field}

To start, we discuss two types of exact solution turning on a non-zero
B-field in four directions.  We work in 25+1 Minkowski space, taking
$g_{\mu\nu}=\eta_{\mu\nu}=(-1,+1,\cdots,+1)$ and
$B_{23,22}=B_{25,24}=b>0$. \foot{We take $\Theta^{ij}$ to be
skew diagonalized.  Unless explicitly stated as in section
2.5, we will set $B_{23,22}=B_{25,24}\,$.  Our results can easily be
extended to the $B_{23,22}\neq B_{25,24}\,$ case with no effect on our conclusions.}  From
\swmap, 
\eqn\mapa{\eqalign{\theta \equiv&~ \Theta^{22,23} = \Theta^{24,25}={1 \over b}~, \cr
G_{\mu\nu}=&~ {\rm diag}(-1,1,\ldots,1,b'^2,b'^2,b'^2,b'^2)~,  \cr
G_{s} =&~ b'^{2}g_{s}~,}}
where $b'\equiv 2\pi\alpha' b$.
 
The soliton solution for $N$ coincident $D21$-branes is given by  
\eqn\NDf{\phi=I-P~,\quad C_{i}=S\ith{ij}x^{j}\bS~,}
where $I$ is the identity operator and $P$ is a rank $N$ projection
operator both acting on $\CH \otimes \CH$.  $S$ is defined as the
operator that satisfies $S\bS=I-P$ and $\bS S=I$.  This solution is an
exact solution by the solution generating technique described in \hkl .  We can
check that this solution gives the right tension. \foot{We integrate the
action to find the tension.  In general, the noncommutative action may
differ from the commutative action by total derivative terms \sw .  In
calculating the tension, these terms do not contribute because for D-brane
soliton solutions the tachyon potential vanishes at infinity.}   Indeed,
with $\sqrt{det(M_{ij})}=b'^{4}P+(b'^{4}+b'^{2})(I-P)$, the evaluation
of the action for this solution gives 
\eqn\NDfE{\eqalign{\CS=&-{c \over {g_{s}b'^{2}}}({2\pi \over
b^{2}})^2b'^{4}V(-1)Tr\,P \int\! d^{22}x \sqrt{-g} \cr = &
-NT_{25}(4\pi^{2}\alpha')^2\int\! d^{22}x \sqrt{-g}  = -N T_{21} \int\!
d^{22}x \sqrt{-g}~,}}
where we have used the relation $T_{p-n}=(4\pi^{2}\alpha')^{\frac{n}{2}}T_{p}$.
Notice that the tension is independent of $b$.  This is a consequence
of the background independence of the action, which would be manifest if we
had used the variables $X^{i}=\Theta^{ij}C_{j}$  (see \refs{\sei,
\deAl}).

Now consider the soliton solution given by
\eqn\NDb{\eqalign{\phi=&(I-P_{1})^{I}I^{\II}~,    \cr
C_{i}=&(S_1\ith{ij}x^{j}\bS_1)^{I}I^{\II}~, \quad  i=24,25~, \cr
C_{i}=&I^{I}(\ith{ij}x^{j})^{\II}~, \quad\quad\quad i=22,23~.}}
The tachyon field solution is exactly that of a single $D23$-brane
solution with $B_{23,22}\ne 0$.  Thus, \NDb\ is the soliton
solution of a single $D23$-brane with tangential B-field in the $x^{22,23}$
directions.  As a check, we take the solution of \NDb\ and evaluate the
action to obtain the expected result
\eqn\NDbe{\CS=-{c \over g_{s}} (4 \pi^2 \alpha')\sqrt{1+b'^2}\int\!
d^{24}x \sqrt{-g}=-T_{23}\sqrt{1+b'^2}\int\! d^{24}x \sqrt{-g}~.}
Here, $\sqrt{\det(
M_{ij})}=b'^3\sqrt{1+b'^2}P_{1}^{I}I^{\II}+(b'^4+b'^2)(I-P_{1})^{I}I^{\II}$.

Notice that \NDb\ can not be generated by the
solution generating technique of \hkl.  If instead we had
$C_{i}=(I-P_1)^{I}(\ith{ij}x^{j})^{\II}$ for $ i=22,23$ in \NDb\ with
the other fields remaining unchanged, then the
solution would be an infinite number of coincident $D21$-branes as it satisfies the
general form of \NDf.  This solution demonstrates the important role the
gauge field plays in determining the interpretation of NC solitonic brane
solutions.  (This issue was also raised in \deAl .)  Note also that such
a distinction between an infinite number of $D21$-branes and a $D23$-brane
with non-zero tangential B-field can not be made in the limit $\alpha' b
\to \infty$ since terms with $C_{i}$ in the action \ncac\ are negligible
in this limit.

\subsec{Perpendicular Brane Configurations and Defect Branes}

Since we have turned on $B_{22,23}$ and $B_{24,25}\,$, it seems plausible
that there may exist solitonic configurations where two $D23$-branes
intersect.  The simplest configuration would be that of one $D23$-brane
with transverse coordinates $x^{22,23}$ intersecting
with another $D23$-brane with transverse coordinates $x^{24,25}$.
Here, each $D23$-brane contains a rank two tangential B-field.  Important to
the NC brane construction is that the tachyon operator $\phi$ must be
of the form $I-P$, where $P$ is a projection operator of finite or
infinite rank.  This form is required to simplify the evaluation of
$V(\phi-I)$ in the action and $V'(\phi-I)$ in the equations of motion.
The solution of a single
$D23$-brane in \NDb\ suggests that a likely candidate for a
configuration of two perpendicularly intersecting $D23$-branes which is also an exact solution is 
\eqn\PBC{\eqalign{\phi=&(I-P_1)^{I}(I-P_{1})^{\II}~, \cr
C_{i}=&(S_{1}\ith{ij}x^{j}\bS_{1})^{I}I^{\II}~, \quad\quad i=24,25~, \cr 
C_{i}=&I^{I}(S_{1}\ith{ij}x^{j}\bS_{1})^{\II}~, \quad\quad i=22,23~.}}
This gives $I-\phi=P_1^II^{\II}+I^IP_1^{\II}-P_1^IP_1^{\II}$, indeed a
projection operator.  However, the presence of the term
$-P_1^IP_1^{\II}$ naively suggests that around the origin,\ \foot{The
functional representation of $P_1$ under Weyl correspondence is
$P_1(x,y)=2e^{-(x^2+y^2)/\theta}$ where $[x,y]=i\theta$.}
where the branes intersect, the configuration might involve an
additional $D21$-brane.  To be more concrete, we evaluate the action for
this configuration.  For the solution of \PBC , we obtain 
\eqn\PBCA{\eqalign{
V(\phi-I)=&V(-1)\{P_1^IP_1^{\II}+P_1^I(I-P_1)^{\II}+(I-P_1)^IP_1^{\II}\}~, \cr
\sqrt{\det(M_{ij})}=&b'^4P_1^IP_1^{\II}+b'^3\sqrt{1+b'^2}(P_1^I(I-P_1)^{\II}+(I-P_1)^IP_1^{\II}) \cr
 &~~~ +(b'^4+b'^2)(I-P_1)^I(I-P_1)^{\II}~,}}
which gives
\eqn\PBCe{\eqalign{\CS=&-{c \over g_s}\left
\{(4\pi^2\alpha')^2+(4\pi^2\alpha')\sqrt{1+b'^2}{2\pi \over b} \left(Tr_{\CH^{I}}(I-P_1)^{I}+Tr_{\CH^{\II}}(I-P_1)^{\II}\right)
\right\} \cr & \qquad\qquad \times \int\! d^{22}x \sqrt{-g} \cr
=&-\left\{T_{21}+T_{23}{{2\pi\sqrt{1+b'^2}} \over b} 
(Tr_{\CH^{I}}(I-P_1)^{I}+Tr_{\CH^{\II}}(I-P_1)^{\II})\right\} \int\! d^{22}x
\sqrt{-g} \cr =& -T_{21}\!\!\int\! d^{22}x
\sqrt{-g}\;-\;2\!\left(T_{23}\sqrt{1+b'^2}\!\int\! d^{24}x
\sqrt{-g}\;-\;T_{21}\frac{\sqrt{1+b'^2}}{b'}\!\int\! d^{22}x\ \right)}}
The action calculation \PBCe\ clearly shows that the configuration
is {\it not}$\,$ that of two perpendicularly intersecting $D23$-branes.
From \PBCe , the configuration is however naturally separated into three
components - a $D21$-brane situated at the origin plus two
perpendicularly-oriented {\it defect} $D23$-branes.   The defect of the
brane is apparent from the $Tr_{\CH}(I-P_1)$ factor.  If the factor was instead
$Tr_{\CH}(I)$, then by the trace-integral correspondence of \intr , we
would have two $D23$-branes in addition to a $D21$-brane.  The presence
of $P_1$ (or in general $P_n$) can be thought of as a ``bubble'' of the
closed string vacuum within the unstable $D$-brane.  This can be seen
most clearly in taking $\alpha'b\to\infty$.  Because of the open string
boundary condition, the $D23$-brane becomes a continuous distribution of
$D21$-branes in the presence of a rank two tangential B-field \sw .   As
observed in \hklm\ and evident from the last equality of \PBCe , a defect
$D23$-brane in this limit is simply a $D23$-brane with one (or in
general n) $D21$-brane subtracted or decayed away.   The decay of
these $D21$-branes into the closed string vacuum forms the bubble.  

Indeed, the defect brane is an exact static solution for any non-zero
value of $b$.  With only a rank two B-field turned on, $B_{25,24}=b>0$,
a defect $D25$-brane solution is simply 
\eqn\dfect{\phi=P_1~,\quad \quad \quad C_i=S_1\ith{ij}x^j\bS_1~ \quad
i=24,25~.} 
This solution gives a nonzero field strength $\hF_{24,25}=-b P_1$.  With
$\sqrt{\det(M_{ij})}=b'^2P_1+b'\sqrt{1+b'^2}(I-P_1)$ and
$V(\phi-I)=V(-1)(I-P_1)$, the evaluation of the action gives
\eqn\dfecte{\eqalign{\CS=&-T_{25}{{2\pi\sqrt{1+b'^2}} \over b} 
Tr_{\CH}(I-P_1) \int\! d^{24}x \sqrt{-g} \cr =&
-\left(T_{25}\sqrt{1+b'^2}\!\int\! d^{26}x
\sqrt{-g}\;-\;T_{23}\frac{\sqrt{1+b'^2}}{b'}\!\int\! d^{24}x\right)\ .}}
In \dfect , both the tachyon field and the resulting field strength
vanish asymptotically.  Thus, far away from the origin, the defect
$D25$-brane has the tension of a $D25$-brane with tangential B-field.  In
general, the defect $Dp$-brane soliton is a non-constant solution with
the boundary condition that it asymptotically approaches the value of a
$Dp$-brane instead of the closed string vacuum.  The defect brane has a
non-constant tension that varies spatially in the noncommutative
directions.\ \foot{Note that the calculation in \dfecte\ does not give the
tension of the defect brane.  In \dfecte\ and also \PBCe , we have simply
integrated the contribution of the ``bubble'' to the action to
demonstrate the difference between defect branes and $D$-branes.}

We can intuitively understand the presence of defect branes in
the decay of a single $D25$-brane into a perpendicular brane
configuration from the ``semiclassical'' viewpoint.  As in Bohr
quantization, we can roughly associate each projection operator with an
area of the noncommutative space.  A $D23$-brane would involve covering
a sheet of noncommutative plane with projection operators such as
$P_1^{I}I^{\II}$.  However, an intersection is an overlap region where
there must be two projection operators present.  But adding two identical
projection operators never results in a projection operator.  Thus, the
requirement that $\phi-I$ be a projection operator prohibits
the construction of two intersecting branes at least as a decay mode of
a single $D25$-brane.  One of the intersecting branes must have a defect.
Here, in our semiclassical description, a defect brane is a $D$-brane
with a ``hole.''  As for the exact solution of \PBC\ consisting of two
perpendicular defect branes, the hole that results from the defect is
filled with a $D21$-brane. 

We leave to the reader straightforward generalizations of
constructing defect branes with ``larger bubbles'' and other perpendicular
configurations by turning on a higher rank B-field.  

\subsec{Perpendicularly Intersecting $D$-Branes}

The above semiclassical argument requires that any intersecting $D$-brane
solutions contain two projection operators covering the same
noncommutative region.  This can be attained by decaying two
D25 branes.  Simply put, with a rank four B-field turned on, two
intersecting $D23$-branes is constructed from two $D25$-branes by letting
each $D25$-brane decay into a $D23$-brane but with different transverse
directions.  The fields now have Chan-Paton indices and are $2 \times 2$
operator-valued  matrices.  The noncommutative action with NC U(2) gauge
symmetry now has an additional trace over the U(2) representation and we
take it to be
\eqn\ncna{\eqalign{
\CS=-{c \over G_s} \int\! d^{26}x \; Tr
&\left \{ V(\phi-1)\sqrt{-\det(G_{ij}-2\pi\alpha'i[C_{i},C_{j}])}\right. \cr
&~~~-\left.{\alpha' \over
2}\sqrt{G}f(\phi-1)G^{ij}[C_{i},\phi][C_{j},\phi]+ \cdots \right \}~,}}

Subtle issues of the non-Abelian DBI action (see, for
example, \atsey ) will not affect our discussion below and will be
ignored.  Using \NDb , the perpendicularly intersecting $D23$-branes
configuration is given by
\eqn\persol{\eqalign{\phi =& \left( \matrix{(I-P_1)^II^{\II} & 0 \cr 0
& I^{I}(I-P_1)^{\II}} \right)~, \cr  
C_{i}=& \left( \matrix{(S_1\ith{ij}x^{j}\bS_1)^{I}I^{\II}  & 0 \cr 0
&(\ith{ij}x^{j})^{I}I^{\II} }\right)~,\quad\quad i=24,25~, \cr
C_{i}=& \left( \matrix{I^{I}(\ith{ij}x^{j})^{\II} & 0 \cr 0
&I^{I}(S_1\ith{ij}x^{j}\bS_1)^{\II}}\right)~,\quad\quad i=22,23~.}}
Since the fields are all diagonal, one easily find that the solution
\persol\ gives the expected tension, two copies of \NDbe .  As a
nontrivial check, one can check for the presence of the ground state tachyonic
fluctuation mode existing at the intersection.  This mode corresponds to
an open string connecting the two branes with both ends at the
origin.  Indeed, the ground state tachyonic mode arises from the
fluctuation of the off-diagonal mode,
\eqn\delphi{\delta\phi =\left( \matrix{0 & \beta |0,0\rangle\langle
0,0| \cr \bbeta |0,0\rangle\langle 0,0| & 0 }\right)~.}
Details of the quadratic fluctuation of the $\beta$ mode are given in
Appendix A.1.  The mass of this mode is found to be 
\eqn\mbeta{\eqalign{m^2_{\beta}=&\frac{1}{\alpha'}\left(\frac{1}{\pi b'}-\frac{\sqrt{1+b'^2}}{b'}\right)
\cr =& \frac{1}{\alpha'}\left[ -1+\frac{1}{\pi b'}
-\frac{1}{2b'^2}+\CO(\frac{1}{b'^3})\right] \quad {\rm as}~b' \to \infty ~. }}
As worked out in Appendix A.2$\,$, the expected mass from the first
quantization calculation is  
\eqn\mfq{\eqalign{m^2=&\frac{1}{\alpha'}\left[-1+(\frac{1}{4}-\frac{1}{\pi^2}tan^{-2}b')\right]
\cr =& \frac{1}{\alpha'}\left[ -1+\frac{1}{\pi b'}-\frac{1}{\pi^2
b'^2}+\CO(\frac{1}{b'^3}) \right] \quad {\rm as}~b' \to \infty ~. }}
Comparing \mbeta\ with \mfq, one sees that we have agreement in the large
$b'$ limit to order $\frac{1}{b'}$.   This is as expected since to obtain
agreement to higher order would require working with higher derivative
terms of the tachyon field in the effective action \ncna.  The exact
form of these higher derivative terms is not needed for constructing
the exact solution but is essential for calculating the exact masses of the
fluctuation modes. 

\subsec{Non-perpendicularly Intersecting Branes} 

So far we have constructed only perpendicularly intersecting $D$-branes.
It is also possible in the operator formalism to obtain
non-perpendicularly intersecting branes.  We will demonstrate this for the case
of two intersecting $D23$-branes.  This again requires working with the
decay of two $D25$-branes.  The solution we seek is of the diagonal form as
in \persol, but now with one $D23$-brane with transverse coordinates $x^{24,25}$ and the other $D23$-brane rotated relative to the first.  Since before the
decay the noncommutative factor $\Theta^{ij}$ is the same on both
branes, our goal is to rotate the $D23$-brane keeping $\Theta^{ij}$ fixed.  

As a rank two tensor, $\Theta^{ij}$, with $i,j=22,\ldots,25\,$,
transforms under the SO(4) rotation in the four noncommutative directions as
\eqn\thtrans{\Theta^{ij}\to R^{i}\!_{k}R^{j}\!_{l}\Theta^{kl}~.}
where $R^{i}\!_{j}$ is an element of SO(4).   With $R=e^{-i\epsilon^{A} J_{A}}$, where $J_{A}$ are the
SO(4) generators in the fundamental representation, the requirement that
$\Theta^{ij}$ is invariant gives the condition
\eqn\thcond{[\epsilon^{A} J_{A}, \Theta]=0~.}
For $\Theta^{22,23}=\theta$ and $\Theta^{24,25}=\theta'$ with $\theta \ne
\theta'$ and other components zero, \thcond\ breaks SO(4)=SU(2)$\times$SU(2) down to
U(1)$\times$U(1).  The U(1)'s correspond to separate rotations in the
$x^{22}\!-\!x^{23}$ and $x^{24}\!-\!x^{25}$ planes.   More
interestingly, for $\theta = \theta'$, the unbroken group is
SU(2)$\times$U(1).  Here, the U(1) generator is proportional to
$\Theta=i\theta({\b 1} \otimes \sigma_2)$ where ${\b 1}$ and
$\sigma_A$ are the 2$\times$2 identity matrix and the Pauli matrices,
respectively.  The SU(2) generators are then 
\eqn\jsu{J_1=-\frac{1}{2}(\sigma_1\otimes\sigma_2)~,\quad
J_2=\frac{1}{2}(\sigma_2\otimes{\b 1})~,\quad J_3=-\frac{1}{2}(\sigma_3\otimes\sigma_2)~.}

As an example, we will work out the rotation under the $J_2$ generator.
This corresponds to a rotation by an angle $\vp$ in both the
$x^{22}\!-\!x^{24}$ and $x^{23}\!-\!x^{25}$ planes, or
\eqn\jexp{[R_{2}]\,^{i}\!_{j}=\left( \matrix{ \cos\,\vp & 0 & -\sin\,\vp
&0 \cr 0 & \cos\,\vp & 0 & -\sin\,\vp \cr \sin\,\vp &0 & \cos\,\vp &
0 \cr 0&\sin\,\vp &0&\cos\,\vp} \right)~.}
The unitary transformation of operators under rotation is determined by
$U^{\dagger}x^{i}U=R^{i}\!_{j}x^{j}$.  Using $[x^i,x^j]=i\Theta^{ij}$, 
the unitary operator associated with $R_2$ is found to be  
\eqn\unitop{U_{2}=exp\left(i\vp\frac{x^{22}x^{25}-x^{23}x^{24}}{\theta}\right)~.}
We now have the necessary ingredients to write down the solution of a
$D23$-brane with transverse coordinates $x^{24,25}$ intersecting
another $D23$-brane situated at an angle $\vp$ in both the $x^{22}\!-\!x^{24}$
and $x^{23}\!-\!x^{25}$ planes from the first.  The solution is
\eqn\angsol{\eqalign{\phi =& \left( \matrix{(I-P_1)^II^{\II} & 0 \cr 0
& U_{2}^{\dagger}(I-P_1)^II^{\II}U_{2}}\right)~, \cr
C_{i}=& \left( \matrix{C'_{i}&0 \cr 0 & [R_{2}]_{i}\!^{j}U_{2}^{\dagger}C'_{j}U_{2} } \right)~,}}
having defined $C'_{i}$ as the gauge field solution $C_{i}$ in \NDb\ for the
single $D23$-brane with rank two tangential B-field.  One can check that
for $\vp=\pi/2$, \angsol\ becomes the solution of the perpendicularly
intersecting $D23$-branes of \persol .  This is most easily done
working with the annihilation operators
\eqn\coor{a_{1}=\frac{x^{24}+ix^{25}}{\sqrt{2\theta}}~,\quad
a_{2}=\frac{x^{22}+ix^{23}}{\sqrt{2\theta}}~,}
and their complex conjugate creation operators.  As an aside, this
operator basis makes explicit that the unbroken
SU(2)$\times$U(1)$\simeq$U(2) rotation when $\theta=\theta'$ is the U(2)
symmetry group of the two-dimensional isotropic oscillator.

\newsec{Superstring}

In the bosonic theory, we have shown that in the presence of at
least a rank four tangential B-field, a $D$-brane can decay into
perpendicular brane configurations.  These brane configurations consist of
what we called defect branes in addition to lower dimensional branes.  
To obtain configurations of intersecting $D$-branes requires the decay of
at least two coincident $D$-branes. 

Our results in the bosonic theory can be extended to type II
string theory.  For the non-BPS $D$-branes, the results are similar to
those of the
bosonic theory with some subtleties due to the tachyon potential having
two degenerate minima (see \refs{\hklm, \hkle}).   For the $D\dD$
system, the situation is more complex in that the form of the noncommutative
effective action for large field strengths is at present not well
understood.\ \foot{Recently, the effective action for the $D\dD$ system
has been calculated in \refs{\kl, \ttu} using boundary string field
theory.}  To sidestep this issue, we analyze the large B-field
limit which allows us to drop all derivative and gauge field terms.  In
this limit, the noncommutative action for a Type IIB $D9$-$\dD$9 system
is simply the tachyon potential,
\eqn\ssac{\CS=-\frac{c}{G_s} \int\! d^{10}x \sqrt{-G} \left\{ V(\bphi\phi -1) +
V(\phi\bphi -1)\right\} }
where the potential is of a Mexican hat-like shape with local maximum at
$V(-1)=1$ and minima at $|\phi|=1$.  From string field theory \witt, the
tachyon solution must satisfy the partial isometry condition,
$\phi\bphi\phi=\phi$.  This is also a sufficient condition for a
solution to satisfy the
equation of motion of \ssac\ with $V(x)=\sum_{n=2}^{\infty}a_nx^n$.  A
solution corresponding to $m\ D7$-branes and $n\ D7$-branes is given by
$\phi=S_n\bS_m$\ \hkl .\ \foot{Here we choose the convention that the RR
charge of a $D7$ brane is $+1$ and is given by the index of $\phi$.}
Notice that this gives
\eqn\phib{\eqalign{ \bphi\phi-I=-P_m~,& \quad \quad V(\bphi\phi
-I)=V(-1)P_m~, \cr \phi\bphi-I=-P_n~,& \quad \quad V(\phi\bphi -I)=
V(-1)P_n~.}}
The projection operator is again utilized to simplify calculations.
Turning on a rank four B-field, the arguments from section 2.3 imply that
perpendicular configurations with defect branes should also exist.  For
example, for $\phi=\bS_1^I\bS_1^{\II}$,
$V(\bphi\phi-I)=V(-1)\left\{P_1^I(I-P_1)^{\II}+(I-P_1)^IP_1^{\II}+P_1^IP_1^{\II}\right\}$.
 To be certain, one should find the solution for the gauge
fields, which we have neglected in the infinite B-field limit, and
calculate the tension.  

One difference in the superstring case is that there exist solutions of
$D7$-branes perpendicularly intersecting $\dD7$-branes for the single
$D9$-$\dD9$ decay.   For a $D7$-brane perpendicularly intersecting a
$\dD7$-brane, the tachyon solution is simply $\phi=S_1^I\bS_1^{\II}$.
In the semiclassical language of section 2.3$\,$, the projection
operators for the $D$-branes and $\dD$-branes exist on separate
noncommutative planes because the potential in \ssac\ consists of two
terms, one associated with $D$-branes and the other with $\dD$-branes.  However, to
obtain a configuration of two intersecting $D7$-branes requires
the decay of two $D9$-$\dD9$ branes.  

\bigskip\medskip\noindent 
{\bf Acknowledgements:}
I am grateful to J. Harvey and E. Martinec for suggesting this topic.
I thank J. Harvey for sharing his insights through numerous
discussions and for commenting on the manuscript.  Also, I thank
P. Kraus for many useful discussions and R. Bao, B. Craps,
F. Larsen, and E. Martinec for helpful conversations.  This work was
supported by NSF grant PHY-9901194 and in part by the Dept. of Education
GAANN fellowship. 

\appendix{A}{Ground State Tachyon Mass at the Intersection of Two
Perpendicularly Intersecting $D23$-Branes with Tangential B-field}

\subsec{Soliton Fluctuation}
We provide details of the calculation for the mass of the ground state
tachyonic fluctuation mode.  The variation is performed on the NC U(2)
action of \ncna\ with respect to the perpendicularly intersecting brane
solution of \persol .  Finding the masses of all the fluctuation modes
is highly nontrivial and will not be discussed here.  The ground state
fluctuation mode comes from the variation of the tachyon field as given
in \delphi.  Explicitly, we have \eqn\delmo{\phi+\delta\phi-I= \left(
\matrix{-P_1^II^{\II} &
\beta P_1^IP_1^{\II} \cr \bbeta P_1^IP_1^{\II}&  -I^{I}P_1^{\II}} \right)~,}
where $\beta$ is the complex fluctuation field of interest with
dependence on $x^a$, the 22 commutative coordinates.  The potential is
assumed to be polynomial and have the form
$V(\phi-1)=\sum_{n=2}^{\infty}a_n(\phi-1)^{n}$.  This gives to $\CO(\beta\bbeta)$
\eqn\varp{V(\phi+\delta\phi-I)=\left( \matrix{ V(-1)P_1^II^{\II}+\frac{1}{2}V''(-1)\beta\bbeta  P_1^IP_1^{\II}
& \cdots \cr \ldots & V(-1)I^IP_1^{\II}+\frac{1}{2}V''(-1)\bbeta\beta P_1^IP_1^{\II} }
\right) ~,}
We have left out the off-diagonal elements because they do not
contribute to the overall trace of the action.  This can be seen in that
multiplying \varp\ is the diagonal matrix
\eqn\fm{\sqrt{\det(M_{ij})} = b'^3\sqrt{1+b'^2}\left(\matrix{P_1^II^{\II}
& 0 \cr 0 & I^IP_1^{\II}}\right)+(b'^4+b'^2)\left(\matrix{(I-P_1)^{I}I^{\II}&
0 \cr 0 & I^I(I-P_1)^{\II}}\right)~.}

As for varying the tachyon kinetic term, it is
convenient to work in the basis of annihilation and creation
operators of \coor.  As is conventional, we denote the corresponding complex
coordinates as $z_{i}$ and $\bz_{i}$, respectively.  The kinetic fluctuation
has the form given by
\eqn\kfluct{\eqalign{\frac{\alpha'}{2}&\sqrt{G} f(\phi-I)G^{\mu\nu}D_{\mu}\dphi
D_{\nu}\dphi \cr  =& \frac{\alpha' b'^4}{2} f(\phi-I) 
\left\{-\frac{b}{b'^2}\sum_{i=1}^{2}\left([C_{z_i},\dphi][C_{\bz_{i}},\dphi]+[C_{\bz_{i}},\dphi][C_{z_i},\dphi]\right)+\partial^a\dphi\partial_a\dphi
 \right\}~, }}
where we have used \mapa\ and noted that $[C_i,\phi]=0$ for the brane
solution \persol.  A straightforward calculation gives
\eqn\sumc{\sum_{i=1}^{2}\left([C_{z_i},\dphi][C_{\bz_{i}},\dphi]+[C_{\bz_{i}},\dphi][C_{z_i},\dphi]\right)=-\left(\matrix{\beta\bbeta
P_1^{I}P_2^{\II} &0 \cr 0 & \bbeta\beta P_2^{I}P_1^{\II}}\right)~,}
and, moreover, 
\eqn\basi{f(\phi-I)=f(-1)\left( \matrix{ P_1^II^{\II} & 0 \cr 0 & I^{I}
P_1^{\II} }\right)~,\quad \partial^a\dphi\partial_a\dphi = \left( \matrix{
\partial^a\beta\partial_a\bbeta P_1^IP_1^{\II} & 0 \cr 0 &
\partial^a\bbeta\partial_a\beta P_1^IP_1^{\II}} \right)~.}

Using \varp-\basi$\,$, the action for $\beta$ up to $\CO(\beta\bbeta)$ is
found to be  
\eqn\abeta{\CS=-\alpha' T_{21} f(-1) \int\! d^{22}x \left\{ \partial^a\bbeta\partial_a\beta +
\left( \frac{2b}{b'^2}+{1 \over \alpha'}\frac{V''(-1)}{f(-1)}\frac{\sqrt{1+b'^2}}{b'}\right)\bbeta\beta\right\}~.}
By definition, $V''(-1)/f(-1)=-1$ since $\phi=0$ corresponds to the
$D25$-brane.  Therefore, the mass of the ground state tachyon is 
\eqn\mbeta{\eqalign{m_{\beta}^2&=\frac{2b}{b'^2}-{1 \over
\alpha'}\frac{\sqrt{1+b'^2}}{b'} \cr &={1 \over \alpha'}\left
( \frac{1}{\pi b'} - \frac{\sqrt{1+b'^2}}{b'} \right)~, }}
where in the second equality of \mbeta , we have used $b'=2\pi\alpha' b$.
It is important to remember that this mass formula has been derived neglecting
contributions to the quadratic fluctuations from the higher derivative
terms of the tachyon field in the effective action \ncac . 

\subsec{Ground State Mass From String Quantization}

For completeness, we work out the ground state mass of an open string
stretching between two perpendicularly intersecting branes in the presence of a
background B-field.  The open string is parametrized by $\tau$ and
$\sigma$ within the region $-\infty < \tau < \infty$ and
$0\le\sigma\le\pi$.  We will quantize the open string that stretches
from the $D23$-brane with transverse coordinates $x^{24,25}$ at
$\sigma=0$ to the $D23$-brane with transverse coordinates $x^{22,23}$ at
$\sigma=\pi$.  For coordinates $x^a$, where $a=0,\ldots ,21$, both open
string endpoints satisfy the Neumann boundary condition
$\partial_{\sigma}X^a=0$.  With non-zero B-field in the remaining
coordinates, the open string boundary conditions are either Dirichlet
(D), $\partial_{\tau}X^i=0$, or mixed (M),
$\eta_{ij}\partial_{\sigma}X^j+2\pi\alpha'B_{ij}\partial_{\tau}X^j=0$, with
\eqn\bcond{\matrix{ & \quad \sigma=0 \quad & \quad \sigma=\pi \cr
X^{22}, X^{23} &  M  & D  \cr
X^{24}, X^{25} &  D  & M }}

As in section 2.4$\,$, we take $B_{23,22}=B_{25,24}=b>0$ and let
$b'=2\pi\alpha'b$.  The boundary conditions are diagonalized in the
linear combinations 
\eqn\compz{Z^1=\frac{1}{\sqrt{2}}\left(X^{22}+iX^{23}\right)~,
\quad\quad Z^2=\frac{1}{\sqrt{2}}\left(X^{24}+iX^{25}\right)~,}
and their complex conjugates.  In these coordinates, \bcond\ becomes
\eqn\bcz{\matrix{
\partial_{\sigma}Z^1+ib'\partial_{\tau}Z^1\arrowvert_{\sigma=0}=0~,
& 
\partial_{\tau}Z^1\arrowvert_{\sigma=\pi}=0~, \cr
\partial_{\tau}Z^2\arrowvert_{\sigma=0}=0~, &
\partial_{\sigma}Z^2+ib'\partial_{\tau}Z^2\arrowvert_{\sigma=\pi}=0~.}}
The mode expansion is thus shifted and given by
\eqn\modex{\eqalign{
Z^1=&i\left(\frac{\alpha'}{2}\right)^{1 \over 2}\sum_{n=-\infty}^{\infty}\left[e^{i\nu_1\pi}e^{-i(n+\nu_1)(\tau+\sigma)}-e^{-i\nu_1\pi}e^{-i(n+\nu_1)(\tau-\sigma)}\right]\frac{\alpha_{n+\nu_1}}{n+\nu_1}~,
\cr Z^2=&i\left(\frac{\alpha'}{2}\right)^{1 \over 2}\sum_{n=-\infty}^{\infty}\left[e^{-i(n+\nu_2)(\tau+\sigma)}-e^{-i(n+\nu_2)(\tau-\sigma)}\right]\frac{\alpha_{n+\nu_2}}{n+\nu_2}~,}}
where both $\nu_1$ and $\nu_2$ have range $[0,1)$ and \bcz\ requires that $\tan\,\nu_1\pi =
\frac{1}{b'}$ and $\tan\, \nu_2\pi =-\frac{1}{b'}$.  The ground state
mass is therefore
\eqn\gma{m^2=\frac{1}{\alpha'}\left[-1 + {1\over
2}\nu_1(1-\nu_1)+{1\over 2}\nu_2(1-\nu_2)\right]~.}
This expression can be simplified by letting $\nu_1={1\over 2}-\epsilon$
and $\nu_2={1\over 2}+\epsilon$.  We thereby obtain
\eqn\gmaa{m^2=\frac{1}{\alpha'}\left[-1
+\left(\frac{1}{4}-\epsilon^2\right)\right]~, \quad \quad \tan\,\epsilon\pi=b'~.}

\listrefs
\end